\documentclass[showpacs,preprintnumbers,aps,twocolumn]{revtex4}
\usepackage{bm}
\usepackage[english]{babel}
\usepackage[latin1]{inputenc}
\usepackage[dvips]{graphicx}
\usepackage{epsfig}
\usepackage{amsmath}
\usepackage{amsfonts}

\begin{document}
\parindent = 0pt
\date{\today} 
\author{M.~Block$^{1}$, E.~ Sch{\"o}ll$^{1}$, and D.~ Drasdo$^{2}$\footnote{Corresponding author: D.Drasdo@warwick.ac.uk}}
\address{$^1$ Institut f{\"u}r Theoretische Physik, Technische Universit{\"a}t Berlin, Berlin, Germany \\
$^{2}$Mathematics Institute and Center for Systems Biology, University of
Warwick, UK, and Interdisciplinary Center for  Bioinformatics,
Univ. of Leipzig, Leipzig, Germany 
}
\title{Classifying the expansion kinetics and critical surface dynamics of growing cell populations}
\begin{abstract}
Based on a cellular automaton model the growth kinetics and the critical 
surface dynamics of cell monolayers is systematically studied by variation of
the cell migration activity, the size of the proliferation 
zone and the cell cycle time distribution over wide ranges.
The model design avoids lattice artifacts and ensures high performance.
The monolayer expansion velocity derived from our simulations  
can be interpreted as a generalization of the velocity relationship for a 
traveling front in the Fisher-Kolmogorov-Petrovskii-Piskounov (FKPP) equation 
that is frequently used to model tumor growth phenomena by continuum models.
The critical surface dynamics corresponds to the Kardar-Parisi-Zhang (KPZ)
universality class for all parameters and model variations studied.
While the velocity agrees quantitatively with experimental observations
by Bru {\em et al}, the critical surface dynamics is in contrast to their interpretation
as generic molecular-beam-epitaxy-like growth.
\end{abstract}
\maketitle
\begin{figure}
  \psfig{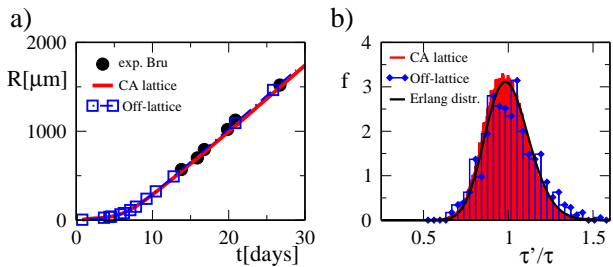}
  \caption{\label{FigLatticeVsOfflattice}
    (Color online) 
    (a) Mean radius $\overline{R}$ of the cell aggregate vs. time $t$.
    Full circles: experimental findings for 
    C6 rat astrocyte glioma cells (\cite{BruAlbSubGarcBru2003:BruEtAl03}).
    (b) Cell cycle time distribution $f(\tau')$ for the off-lattice model and the 
    CA growth model in comparison with the Erlang distribution.
    ($m=60$, $\Delta L=9$, $\phi = 0$)
  }
\end{figure}
\begin{figure}
  \psfig{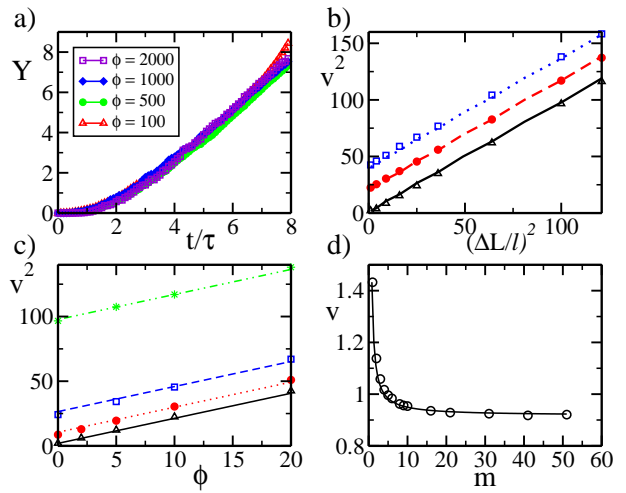}
  \caption{\label{FigExpansionVel}
    (Color online) 
    (a) $Y=R_{gyr}^2/(\phi+1/\tau_\mathrm{eff})$ vs. $t/\tau$ for $m=0$, $\Delta L=1$ and 
    different values for $\phi$. 
    (b-d): Growth in the linear expansion regime ($N\sim 10^5$).
    (b) Square of expansion velocity, $v^2$, vs. square of the
    proliferation zone, $\Delta L^2$
    (triangles: $\phi=0$, circles: $\phi=10$, squares: $\phi=20$; $m=0$).
    (c) $v^2$ vs. $\phi$ (triangles: $\Delta L = 1$, circles:
    $\Delta L = 3$, squares:  $\Delta L = 6$, stars:  $\Delta L = 10$; $m=0$).
    (d) $v$ vs. $m$ ($\Delta L = 1$, $\phi=0$). 
    The lines are fits using eqn. (\ref{EqnVelocity}).
  }
\end{figure}

Model simulations of tumor growth and therapy have attracted 
wide interest \cite{GatenbyMaini2003}-\cite{DrasdoHoehme2005}.
An important issue to which models can contribute is the
classification of the tumor 
growth pattern by generic mechanisms at the level of the individual cell 
actions (migration, division etc.).
These actions subsume the effect of the molecular inter-and intra-cellular 
regulation. 
The models can serve to identify those cell activities that would result in a 
maximal inhibition of multi-cellular growth and invasion, and thereby help to
identify possible molecular drug targets.
Bru {\em et al} \cite{BruAlbSubGarcBru2003:BruEtAl03} analyzed the growth kinetics and critical
surface dynamics of many different tumors in-vitro and in-vivo.
They quantified the dynamics of the tumor surface by three {\em critical exponents}
used to classify crystal growth phenomena into universality classes 
\cite{BarabasiStanley1995:BaSt95}.
They found a generic linear growth phase of in-vitro growing cell lines 
for large cell populations and a molecular-beam-epitaxy (MBE)-like dynamics 
of the tumor surface both in-vitro and in-vivo.
They proposed a tumor therapy based on these findings \cite{BruEtAl2004}.
\\
In this letter we analyze a class of cellular automaton (CA) tumor growth models 
on an irregular lattice by extensive computer simulations. 
CA tumor growth models enjoy wide interest \cite{MoreiraDeutsch2002} since
they permit to represent each cell individually at moderate
computational expense.
In our model cells can divide, push neighbor cells and migrate.
The choice of the model rules is guided by comparison with an off-lattice model.
By using the irregular lattice we ensure isotropy and homogeneity of space, 
and cell sizes that are sharply peaked around a prescribed average value.
Both the expansion speed and the spatial pattern formed differ from results
on a periodic lattice.
We systematically analyze our growth model with respect to the
hopping rate, proliferation depth and dispersion of the cell cycle time
distribution and show that the expansion dynamics can be mapped onto the
functional form of the traveling wave velocity of the 
Fisher-Kolmogorov-Petrovskii-Piskounov (FKPP) equation \cite{Murray1989}.
The model reproduces the monolayer expansion kinetics experimentally found by Bru
\cite{BruAlbSubGarcBru2003:BruEtAl03}.
The critical surface growth dynamics suggests a Kadar-Parisi-Zhang (KPZ)-like 
\cite{KPZ1986} behavior over a wide
range of parameters and for varying cell migration
mechanisms, supporting the critical comment by Buceta and Galeano
\cite{BucetaGaleano2005} on the conjecture by Bru et. al.
\cite{BruAlbSubGarcBru2003:BruEtAl03}.
Our findings comply with the results in the classical Eden model \cite{Moro2001}.\\
\begin{figure}
  \psfig{figure=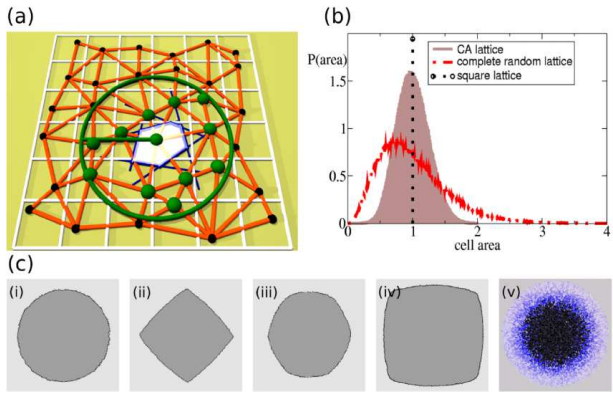,width=0.45\textwidth}
  \caption{\label{Fig:lattice_basics}
    (Color online)
    (a) Construction of the CA lattice: 
    One point (black, green) is placed in every square of a square lattice 
    at a random position $\underline{r}_i$.
    A Voronoi tesselation is constructed form these points such that each cell
    consists of all points in space that are closer to the lattice 
    point $\underline{r}_i$ than to any other $\underline{r}_k$.
    The shape of a biological cell (white) is identified with the corresponding 
    Voronoi polygon
    (blue lines). Polygons that share a common edge are defined as neighboring
    and connected by red lines (Delauney triangulation).
    (b) Probability density distribution of the cell area for the CA-lattice in (a) (brown)
    and for a random initial distribution of points (red).
    (c) Cell cluster morphology for $m=10^4$, $\Delta L=1$ on (i) the CA lattice
    in 
    (a), (ii) square-, (iii) hexagonal lattice, (iv) lattice with 
    Moore neighborhood (nearest neighbors along the axes and the diagonals),
    (v) off-lattice cluster \cite{DrasdoHoehme2005,Drasdo2005}.
  }
\end{figure}
Our model is based upon the following assumptions:\\
$[R1]$ \textit{Lattice generation}:
Starting from a regular square lattice with spacing $l$, an irregular lattice 
$\underline{r}_i$ is generated by Delauney triangulation. 
A biological cell is represented as shown in Fig.\ref{Fig:lattice_basics}(a) (white).
\\
$[R2]$ \textit{Exclusion principle}:
Each lattice site can be occupied by at most one single cell. \\
$[R3]$ \textit{Cycle time}:
The cell cycle time $\tau'$ is Erlang distributed (with a parameter $m$):
\begin{eqnarray}
f(\tau')=\lambda_m\frac{(\lambda_m\tau')^{m-1}}{(m-1)!}exp\{-\lambda_m\tau'\}
\label{EqnErlangdistr}
\end{eqnarray}
with $\lambda_m=m$ such that $\langle\tau'\rangle\equiv\tau=1$.\\
$[R4]$ \textit{Proliferation depth}:
A cell can divide if and only if there is at
least one free neighbor site within a circle of radius $\Delta L$ around
the dividing cell (Fig. \ref{Fig:lattice_basics} (a), green).\\ 
$[R5]$ \textit{Cell migration}:
We consider three alternative migration rules:
R5(i) A cell moves with rate $\phi$ to a free neighbor site, irrespectively
of the number of neighbor cells before and after its move.
This rule corresponds to the case of no cell-cell adhesion.
R5(ii) Cells move with rate $\phi$ if by this move the cell is not isolated.
R5(iii) Cells move with a rate $\phi\exp\{-\Delta E/F_T\}$ with
$\Delta E=E(t+\Delta t)-E(t)$, where $\Delta t$ is the time step,
$E(t)$ is the total interaction energy of
the multi-cellular configuration, $F_T\sim 10^{-16}J$ is a ''metabolic''
energy \cite{BeysensForgacsGlazier2000:BFG00}, 
$\Delta E/F_T \sim {\cal O}(1)-{\cal O}(10)$ \cite{DrasdoHoehme2005}.
This induces migration towards locations with a larger number of neighbor
cells.
\\
By [R1] we generate an unstructured lattice with a symmetric cell area
distribution sharply peaked around its average $A = l^2$ 
(see Fig.\ref{Fig:lattice_basics} (a),(b)).
[R3] considers that experiments indicate a $\Gamma$-like distribution of 
the cell cycle controlled by cell cycle check points \cite{AlbertsEtAl2002}. 
[R4] takes into regard that the growth speed of tumors is usually incompatible with the assumption that
only cells at the border are able to divide (as in the Eden model 
\cite{Eden1961:ME61}, see \cite{DrasdoHoehme2005}).
Therefore we assume that a dividing cell is able to trigger the migration of at most 
$k$ neighbor cells into the direction of minimum mechanical stress (see 
Fig.\ref{Fig:lattice_basics} (a)).
If a cell divides, one of its daughter cells is placed at the original
position, the other cell is placed next to it and
the local cell configuration is shifted and re-arranged along the line that
connects the dividing cell with the closed free lattice site within a circle of
radius $\Delta L$ such that the latter  is now
occupied (see Fig.\ref{Fig:lattice_basics} (a)).
This algorithm mimics a realistic re-arrangement process that may occur from
active cell migration as a response to a mechanical stimulus, cf. Ref. \cite{KansalEtAl2000}.
Isolated cells perform a random-walk-like motion (e.g.
\cite{Schienbein1994:SchFrGru94}).
We consider different migration rules R5(i)-(iii) to comprise a class of potential 
models with biologically realistic behavior.\\
The model parameters are the average cell cycle time $\tau$ and its distribution 
$f(\tau')$ controlled by the parameter $m$, 
the migration rate $\phi$, the proliferation 
depth $\Delta L$, and, in case of an  
energy-activated migration rule, the energy $E$.
Programmed cell death can easily be integrated \cite{apoptosis} but is 
omitted here.
Rules [R1-R5] can be formalized by the master equation
\begin{eqnarray}
\partial_t p(Z,t)=\sum_{Z'\rightarrow Z}W_{Z'\rightarrow Z}p(Z',t)- W_{Z\rightarrow Z'}p(Z,t).
\label{EqnMaster}
\end{eqnarray}
Here $p(Z,t)$ denotes the multivariate probability to find the cells
in configuration $Z$ and $W(Z'\rightarrow Z)$
denotes the transition rate from configuration $Z'$
to configuration $Z$.
A configuration $Z=\{..., x_{i-1}, x_i, x_{i+1}, ...\}$ consists of
local variables $x_{i}=\{0,1\}$ with $x_i=0$ if lattice site $i$ is empty,
and $x_i=1$ if it is occupied by a cell.
For the simulation we use the Gillespie algorithm
\cite{Gillespie1976}, i.e,
the time-step of the event-based simulation is a random number given by $\Delta t =
-\frac{1}{W_{Z}}ln(1- \xi)$.
Here, $\xi$ is a random number equidistributed in $[0,1)$,
$W_Z=\sum_{Z'} W_{Z'\rightarrow Z}$
is the sum of all possible events which may occur at time $t$.
Here we assume that the rate at which a cell changes its state by 
a hop, a progress in the cell cycle, or a division is independent of the
number of accessible states as long as at least one state, that is, 
one free adjacent lattice site in case of a hop and one free site 
within a circle of radius $\Delta L$ in case of a division, is accessible.
This may be justified by noting that cells - in contrast to physical particles -
are able to sense their environment and therefore the direction into which they 
can move.\\
We analyze the growth kinetics by the cell population size $N(t)$ 
(number of cells at time $t$) 
and the radius of gyration $R_{gyr}(t) = \sqrt{\frac{1}{N}\sum_{i=1}^N
(\underline{r}_i-\underline{R}_0)^2}$.
Here $\underline{R}_0=\frac{1}{N}\sum_{i=1}^N\underline{r}_i$ is the position of the
center of mass.
For a compact circular cell aggregate (in $d=2$ dimensions), $R_{gyr}$ is
related to the mean radius $\overline{R}(t)=
\frac{1}{2\pi}\int_{0}^{2\pi}R(\varphi,t)d\varphi$ 
(polar angle $\varphi$)
of the aggregate by $\overline{R} = R_{gyr} \sqrt{2}$.\\
To interpret the rules and parameters of the CA model in terms of growth
mechanisms we compare it with the stochastic single-cell-based off-lattice 
growth model in Ref. \cite{DrasdoHoehme2005} 
(Fig. \ref{FigLatticeVsOfflattice}).
In this model cell motion contains an active random component and a component 
triggered by mechanical forces between cells, and between cells and the substrate \cite{ChuEtAl2005}.
During cell division the cell gradually deforms and divides into two daughter
cells as long as the degree of deformation and compression is not too large.
As illustrated in Fig. \ref{FigLatticeVsOfflattice} the lattice model is able 
to capture the behavior of the off-lattice model and agrees with the 
experimental findings in Refs. \cite{BruAlbSubGarcBru2003:BruEtAl03} provided 
the parameters $\Delta L$, $\phi$, $\tau$, $m$ are chosen properly. 
$\Delta L$ controls the effective thickness of the proliferative rim;
in the off-lattice model it depends on the mechanisms that control the 
proliferation by contact inhibition, on the material 
properties of the cell (the Young modulus, the Poisson number etc.), and 
on the ability of a cells to move in response to a mechanical stimulus 
\cite{DrasdoHoehme2005}.\\
At large $m$ the tumor border becomes smoother and the tumor shape reflects the 
symmetry of the underlying lattice (Fig. \ref{Fig:lattice_basics} 
(c)(ii-iv)); this effect is known as {\em noise reduction} 
\cite{BatchelorHenry1981}.
Such lattice-induced asymmetries could significantly disturb the analysis of 
the surface growth dynamics in circular geometries.
We have chosen a Voronoi tesselation, in which such artifacts do not occur
(Fig. \ref{Fig:lattice_basics} (a),(c)(i)).
Fig.~\ref{FigExpansionVel} shows a systematic study of the growth kinetics for 
free hopping (Rule R5(i)).
All quantities are plotted in multiples of $\tau$ and $l$, which
are the reference time and length scale, respectively.
Initially, the cell population size grows exponentially fast with 
$N(t)=N(0) exp(t/\tau_\mathrm{eff})$ where
$\tau_\mathrm{eff}^{-1}=(2^{1/m}-1)m\tau^{-1}$ \cite{Drasdo2005}.
The duration of the initial phase increases with $\Delta L$ and $\phi$.
The growth law for the diameter depends on $\phi$.
If $\phi=0$, the initial expansion of the diameter is exponentially fast, too.
If $\phi>0$, cells initially detach from the main cluster and the diameter 
grows diffusively, with 
$L\equiv 2\sqrt{2}R_{gyr}\propto \sqrt{A(\phi+1/\tau_\mathrm{eff})t}$ 
where $A\approx 1.2$ is a lattice-dependent fit constant 
(Fig.~\ref{FigExpansionVel}(a)).
For $t/\tau \leq  2$, $R_{gyr}\propto t$ (Fig.~\ref{FigExpansionVel}(a)).
This regime disappears for $N(0)\gg 1$ (see \cite{Drasdo2005}).
As soon as cells in the interior of the aggregate are incapable of further
division the exponential growth crosses over to a linear expansion phase.
Fig.~\ref{FigExpansionVel} shows $v^2$ vs. (b) $(\Delta L)^2$, (c) $\phi$, and 
(d) $m$ for large $N$ ($N\sim 10^5$ cells).
The model can explain the experimentally observed velocity-range in Ref. 
\cite{BruAlbSubGarcBru2003:BruEtAl03}.
As $t\rightarrow\infty$, $L=v(m,\phi,\Delta L)t$ with
\begin{eqnarray}
v^2\approx B^2([\Delta L'(\Delta L)]^2/\tau_\mathrm{eff}^2 + \phi/\tau_\mathrm{eff}), 
\label{EqnVelocity}
\end{eqnarray}
$B\approx 1.4$ (lines in Fig. \ref{FigExpansionVel}b-c).
$\Delta L'(\Delta L)$ ($\approx 1+0.6(\Delta L-1)$) 
results from the average over all permutations to pick 
boundary cells within a layer of thickness $\Delta L$. 
For $\Delta L/\tau_\mathrm{eff}\ll \sqrt{\phi/\tau_\mathrm{eff}}$ eqn. (\ref{EqnVelocity}) 
has the same form as for the FKPP equation. (e.g. \cite{Moro2001}).\\
\begin{figure}
  \psfig{figure=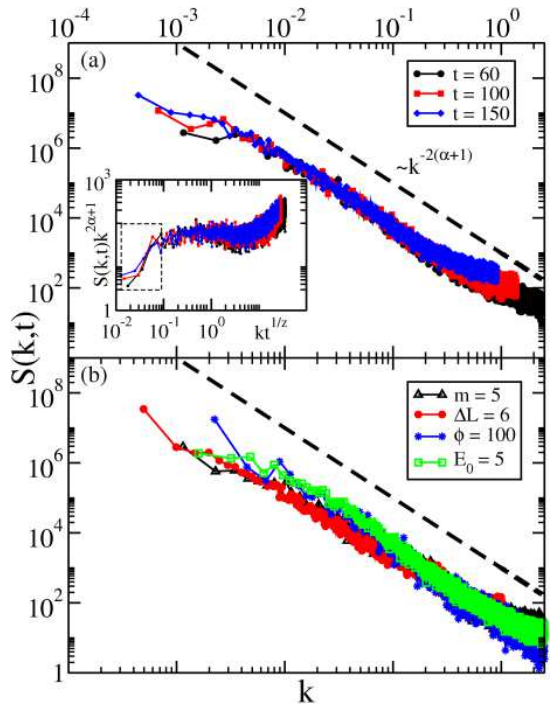,width=0.4\textwidth}
  \caption{\label{FigDynStrFct}
    (Color online) 
    (a) Dynamic structure function $S(k,t)$ vs. $k$ for [R5(i)], $\Delta L=0$, 
    $\phi=0$, $m=0$.
    Inset: rescaled function $S(k,t)k^{2\alpha +1}$ vs. $kt^{1/z}$ ($\alpha=0.5$, 
    $z=3/2$).
    (b) $S(k,t)$ vs. $k$ for four alternative parameter sets: 
    (A) triangles: $m = 5$ ($\Delta L=0$, $\phi=0$),
    (B) circles: $\Delta L=6$ ($m=0$, $\phi=0$),
    (C) stars: R5(ii) $\phi=100$ ($m=0$, $\Delta L=0$), 
    (D) squares: R5(iii) $\Delta E/F_T = E_0+n\cdot E_B$ with $E_B =10$, 
    $n$ neighbors, $E_0=5$ surface binding ($m=0$, $\Delta L=0$)
    \cite{comment_strfnc}.
    The dashed lines are guides to the eye showing $\alpha=0.5$.
  }
\end{figure}
Next, to determine the universality class we determine the roughness exponent
$\alpha$ and the dynamic exponent $z$ from the dynamic structure function
$S(k,t) =\langle R(k,t)R(-k,t)\rangle$ where $R(k,t)$ is the Fourier transform
of the local radius $R(s,t)$ and $\langle ...\rangle$ denotes the average over
different realizations of the growth process (e.g.
\cite{RamascoLopezRodriguez2000:RaLoRo00}).
Here $s$ is the arclength as in Ref. 
\cite{BruAlbSubGarcBru2003:BruEtAl03}.
The third exponent, the growth exponent $\beta$, can be obtained from the 
scaling relation $\beta=\alpha/z$.
In test simulations comparing constant angle segments $\Delta\varphi$ with
constant arclength intervals $\Delta s$ we did not find  noteworthy 
differences.
For self-affine surfaces in absence of any critical length-scale the dynamic
structure function has the Family-Vicsek scaling form
\cite{FamilyVicsek1985:FaVi85}:
\begin{eqnarray}
  S(k,t)&=&k^{-(2\alpha+1)}s(kt^{1/z})\\
s(u=kt^{1/z}) &=&
\left\{
\begin{array}{ccc}
const. & if & u\gg 1 \\
u^{-(2\alpha+1)} & if & u \ll 1.
\end{array}
\right.
\end{eqnarray}
At $u=1$ a crossover occurs.
For $u\gg 1$ curves measured at different times collapse onto a
single line; at $u\ll 1$ they split.
We have calculated $S(k,t)$ for rules R5(i) and $\phi\geq 0$, R5(ii) and
R5(iii) (Fig.~\ref{FigDynStrFct}). 
The final cell population size was of ${\cal O}(10^5)$ cells which is the
typical size of the cell populations in Ref. \cite{BruAlbSubGarcBru2003:BruEtAl03}.
All these results suggest KPZ-like dynamics with $\alpha=1/2$,
$z=3/2$  and $\beta=1/3$
rather than the MBE universality class, i.e., critical exponents $\alpha=3/2$, 
$z=4$ and $\beta=3/8$
inferred in \cite{BruAlbSubGarcBru2003:BruEtAl03}. 
The parameter range of $\phi\in [0,100)$ captures most cell lines studied
in Ref. \cite{BruAlbSubGarcBru2003:BruEtAl03} (for $l=10\mu m$, $\tau = 24h$, 
$\phi = 100$ corresponds to a diffusion constant of $D=10^{-10}cm^2/s$). \\\\
In conclusion we have analyzed the expansion kinetics and critical surface
dynamics of two-dimensional cell aggregates by extensive computer simulations within a CA
model which avoids artifacts from the symmetry of regular lattices.
The growth scenarios are compatible with experimental observations.
The asymptotic expansion velocity has a form that is reminiscent 
of the front velocity of the FKPP equation.
The same expansion velocity can be obtained for different combinations of
the migration and division activities of the cell and of the cycle time
distribution. 
Recently, mathematical models based on the FKPP equation were used to predict 
the distribution of tumor cells for high-grade glioma in regions which are 
below the detection threshold of medical image techniques 
\cite{SwansonAlvordMurray2000:SwAlMu00}.
We believe such predictions must fail since the FKPP equation lacks some
important parameters such as the proliferation depth which is why it is not 
sensitive to relative contributions of the proliferation depth and free 
migration.
We observed in our simulations that these relative contributions in fact
determine the cell density profile at the tumor-medium interface: the
larger the fraction of free migration is, the wider is the front profile
even if the average expansion velocity is constant.\\
The critical surface dynamics found in our simulations does not comply
with the interpretation of experimental observations by 
Bru et. al.  \cite{BruAlbSubGarcBru2003:BruEtAl03} even for the migration 
mechanism they suggested (R5(iii)). 
We propose to re-analyze the corresponding experiments and track the paths of 
marked cells.
\\\\
Support within Sfb 296 (MB) and by DFG grant
BIZ 6-1/1 (DD) is acknowledged.


\begin{thebibliography}{10}
\bibitem{comment_strfnc}
We have also tested $S(k,t)$ vs. $k$ for R5(i), $\Delta L=0$, $\phi=10$, $m=0$
and found $\alpha \simeq 0.5$.
\bibitem{apoptosis}
We mainly found a rescaling of the 
proliferation rate $\tau^{-1}$ to $\tau^{-1}-\gamma$ ($\gamma$ is the rate of 
programmed cell death).
\end{thebibliography}
\end{document}